\documentstyle[preprint,aps]{revtex}
\begin{document}
\preprint{SNUTP-93-41}
\draft
\title{ Phase Space Structure of \\
  Non-Abelian Chern-Simons Particles}

\author{Myung Ho Kim\cite{mhkim}}
\address{ Department of Mathematics,
 Sung Kyun Kwan University,
 Suwon 440-746, KOREA }
\author{Phillial Oh\cite{poh}}
\address{Department of Physics,
Sung Kyun Kwan University,
Suwon 440-746,  KOREA}

\maketitle

\begin{abstract}
We investigate the classical phase space structure of $N$ $SU(n+1)$
non-Abelian Chern-Simons
(NACS) particles by first constructing the product space of
 associated  $SU(n+1)$ bundle with ${\bf CP}^n$ as the fiber.
We calculate the Poisson bracket using the symplectic structure
on the associated bundle and find that the minimal
substitution in the presence of external gauge fields is equivalent to
the modification of symplectic structure by the addition of field
strength two form.
Then, we take a direct product of  the associated bundle by  the space of
all connections and choose a specific connection by the condition
of vanishing momentum map corresponding to the gauge transformation,
thus recovering the quantum mechanical model of NACS
particles in Ref.\cite{lo1}.
\end{abstract}


\narrowtext

\newpage
\def\theequation{\arabic{section}.\arabic{equation}}
\section{Introduction}
\setcounter{equation}{0}
Exotic statistics\cite{wilz3} in two spatial dimensions has attracted
much attention recently\cite{fort,ieng}. It
has relevance to various areas of physics and  in particular
could be realized in condensed matter physics\cite{pran}.
Particles exhibiting exotic statistics and carrying  anomalous
spin, called anyons, can be described as  particles carrying both
 charge and magnetic flux and a simple model for them can be
 constructed by coupling their charges minimally with
the Abelian Chern-Simons gauge field.
The notion of
anyon can be generalized by introducing a non-Abelian gauge group, {\it i.e.},
particles which carry non-Abelian charges and interact with each other
 through  the non-Abelian Chern-Simons theory\cite{dese,witt}.
 One can also construct the classical action for these non-Abelian
Chern-Simons(NACS) particles
by introducing isospin degree of freedom and coupling isospin charge
with the non-Abelian Chern-Simons gauge fields \cite{lo1,bala}.
The resulting  quantum mechanical model of $SU(2)$
NACS particles\cite{lo1,verl}
 indicates that they also carry anomalous spins and satisfy the
 non-Abelian braid statistics\cite{froh}. In Ref.\cite{oh}, a model of
 $SU(n+1)$ NACS particles with arbitrary $n\geq 1$ has been achieved
by considering the isospin degree of freedom defined on ${\bf CP}^n$
manifold. It can be generalized to arbitrary group
 with invariant nonsingular metric\cite{bak}.
 Also, equivalent  second-quantized description  of NACS particles
has been formulated\cite{bak,kiml}.

In this paper  we investigate the model of Ref.\cite{lo1,oh} from a more
geometrical point of view and explore the symplectic structure of the phase
space of the NACS particles.
In order to do this, we first  identify the phase space of
isospin particles in the presence of external gauge fields. Naively, one
 could try to construct the phase space by introducing the tangent bundle
 $T^*G$ of group manifold $G=SU(n+1)$ for the isospin degree of freedom
and taking direct product of this by the canonical phase space.
Then, to account for the external gauge field, one could consider
associated bundle
with $T^*G$ as the fiber\cite{wein}.
 Although this procedure works eventually,
one has to go through some extra reduction procedure to achieve the goal.
More economical way has been known for some time\cite{ster}
 and it is to use
associated bundle with coadjoint orbit\cite{kril} as the fiber.
It can be shown that
the symplectic structure defined on this associated bundle with
 a given connection
is automatically gauge invariant and
 can be used to write down the equations
of motions of a classical particle in the presence of gauge fields.
 The resulting
equations of motion are the well-known Wong's equations\cite{wong}.

To make a connection with the quantum mechanical
model of NACS particles,
we have to work with a specific connection.
This connection, for example, was
determined as a solution of the Gauss constraints in the holomorphic gauge
in Ref.\cite{lo1,oh}. In this more geometrical approach,
this can be achieved by introducing
the space of all connections and take a direct product of this
by the associated bundle constructed previously.
Then, gauge transformation can be defined on the product space and the
corresponding momentum map can be constructed. The condition of
vanishing of this momentum map yields the Gauss law constraints which
can be solved as before.

In this paper, we elaborate on the above procedure using the associated
bundle with ${\bf CP}^n$ as the fiber.
In Sec. 2, we start with a brief review on symplectic geometry and
momentum map which are essential ingredients of phase space structure.
In Sec. 3, we prove that the associated bundle is
 isomorphic to direct product
of  canonical tangent bundle by ${\bf CP}^n$ manifold and evaluate
 the Poisson
bracket between the dynamical variables using the symplectic structure .
We show that the minimal substitution
is equivalent to the modification of symplectic structure
 by the addition of field
strength two form.
In. Sec. 4, we construct the momentum map corresponding to the
gauge transformation and use the
solution of Gauss constraints in holomorphic gauge to recover the
model of Ref.\cite{lo1,oh}.

\section{Momentum Map for ${\bf CP^n}$}
\setcounter{equation}{0}
We start with a brief summary of symplectic geometry\cite{guil}.
Consider a phase space ${\cal O}$
 with symplectic structure $\Omega_{AB}$.
Using the symplectic structure, we define Poisson bracket\cite{bere,fa}:
\begin{equation}
\{F,H\}=\Omega^{AB}\partial_AF\partial_BH \label{poi}
\end{equation}
where $F,H\in C^\infty({\cal O})$ and $\Omega^{AB}$ is the inverse
matrix of  $\Omega_{AB}$.
We define for a given $F$, the corresponding vector field  $\zeta_F$ by
\begin{equation}
 \zeta_F\rfloor\Omega+ dF=0.
\end{equation}
The vector field $\zeta_F$ preserves
 the symplectic structure  ${\cal L}_{ \zeta_F}
\Omega=0$ and $\zeta_F$ defines one-parameter family of canonical
transformation of ${\cal O}$. $\zeta_F$ are called Hamiltonian vector field
generated by $F$ and the set of all Hamiltonian vector fields on ${\cal O}$ is
denoted by $HV({\cal O})$.
One can show that the Poisson bracket defines
 a Lie algebra homomorphism
of  $C^\infty({\cal O})$ onto $HV({\cal O})$:
\begin{equation}
[\zeta_F,\zeta_H]=\zeta_{\{F,H\}}
\end{equation}

Let us assume that ${\cal O}$ is a Hamiltonian $G$-space.
This means that ${\cal O}$ is a symplectic manifold  and the
$G$ group action
$L_g:{\cal O}\rightarrow {\cal O}$ defined by
$f^g=gf$($g\in G,f\in {\cal O}$) is a
 symplectomorphism:
\begin{equation}
L_g^*\Omega=\Omega.
\end{equation}
The case in which ${\cal O}$ is the coadjoint orbit\cite{kril}
of group $G$ is most interesting.
The mapping $\sigma:{\cal G}\rightarrow HV({\cal O})$
 given by $\sigma(\zeta)f=
d/dt{\big\vert}_{t=0}(\exp t\zeta)\circ f$ where $\zeta\in {\cal G}$
is a Lie algebra homomorphism:
$\sigma([\zeta_1,\zeta_2])=[\sigma(\zeta_1),\sigma(\zeta_2)]$.
We assume that there is a lifting $\tilde\sigma:
{\cal G}\rightarrow C^\infty ({\cal O})$
such that the Lie algebra structure is given by the Poisson bracket
on ${\cal O}$:$\tilde\sigma([\zeta_1,\zeta_2])=\{\tilde\sigma(\zeta_1),
\tilde\sigma(\zeta_2)\}$. Thus, to each $\zeta\in{\cal G}$,
 we get a function $\tilde\sigma
(\zeta)\equiv F_\zeta$ and a Hamiltonian
 vector fields $\zeta_F$ of ${\cal O}$
so that
\begin{equation}
\zeta_F\rfloor \Omega+dF_\zeta=0.\label{moment}
\end{equation}
We define the momentum mapping $Q:{\cal O}\rightarrow {\cal G}^*$ where
\begin{equation}
\langle \zeta, Q(f)\rangle=F_\zeta(f), \quad f\in {\cal O}.
\end{equation}
Here ${\cal G}^*$ denotes the dual algebra of ${\cal G}$
 and $\langle,\rangle$
denotes the pairing between ${\cal G}^*$ and ${\cal G}$.

The momentum map for $SU(n+1)$ action on ${\bf CP}^n$\cite{grif} can be
calculated according to the above definition.
Let $u_0, u_1,\cdots, u_n$ be coordinates on ${\bf C}^{n+1}$ and
consider the $SU(n+1)$ group action on ${\bf C}^{n+1}$ given by
\begin{equation}
u^g= gu, g=\exp(-i\theta^a T^a)\in SU(n+1)
\end{equation}
where $u^T=(u_0,\cdots,u_n)$ and the $T^a$'s are the generators of the
 Lie algebra $su(n+1)$:
\begin{equation}
[T^a, T^b]=if^{abc}T^c
\end{equation}
We use the normalization $Tr(T^aT^b)=(1/2)\delta_{ab}$.
Now we introduce the coordinates ${\xi_p=u_p/u_0} (p,q=1,\cdots,n)$ on
the open set $U_0=(u_0\neq 0)$ in ${\bf CP}^{n}$.

To calculate the vector field generated by the generator $T^a$
on ${\bf CP}^{n}$, we  note that
\begin{equation}
\zeta_p=\frac{d}{dt}{\Big\vert}_{t=0}\frac{(\exp\{-itT^a\})_{ps} u_s}
{(\exp\{-itT^a\})_{os} u_s}  \ \ \ \ \ \ \ \ \ (s,t=0,1,\cdots,n)
\end{equation}
\[=-i\left[(T^a)_{po}+(T^a)_{pq}\xi_q-(T^a)_{00}\xi_p-   (T^a)_{0q}\xi_q\xi_p
\right].\]
Hence the vector field generated by $T^a$ is given by
\begin{equation}
\zeta_a=-i\left[(T^a)_{po}+
(T^a)_{pq}\xi_q-(T^a)_{00}\xi_p-(T^a)_{0q}\xi_q\xi_p
\right]\frac{\partial}{\partial \xi_p} + (c.c).
\end{equation}
The above vector field is in fact a Hamiltonian vector field
since the unitary group $SU(n+1)$ acts transitively on ${\bf CP}^{n}$
and leaves the following symplectic two form $\Omega$ on ${\bf CP}^{n}$
invariant:
\begin{equation}
\Omega=2iJ\left[\frac{d{\bar\xi}\wedge d\xi}{1+
\mid\xi\mid^2}-\frac{( \xi d{\bar\xi})\wedge({\bar\xi}d\xi)}{(1
+\mid\xi\mid^2)^2}\right],\ \ \ \mid\xi\mid^2=\sum_p\mid\xi_p\mid^2.
\label{omeg}
\end{equation}
Thus  the momentum map function $Q^a\equiv F_{T^a}$ on ${\bf CP}^{n}$
associated with the Hamiltonian vector field $\zeta_a$  defined by
Eq.(\ref{moment}) satisfy the following:
\begin{equation}
dQ^a=2J(d\xi\cdot \frac{\partial}{\partial \xi}+
d\bar\xi\cdot \frac{\partial}{\partial \bar\xi})\left[\frac{
(T^a)_{00}+(T^a)_{0q}\xi_q+\bar\xi_p(T^a)_{po}+\bar\xi_p(T^a)_{pq}\xi_q}
{1+\mid\xi\mid^2}\right].
\end{equation}
One can show that $Q^a$ can be written  compactly as
 follows(up to a constant
which we neglect):
\begin{equation}
Q^a=2J\sum_{s,t=0}^{N}{\bar u}_s(T^a)_{st}u_t{\Big\vert}
_{u_0=\frac{1}
{\sqrt{1+\mid\xi\mid^2}}, u_p=u_0\xi_p}.\label{def2}
\end{equation}
For example, for $SU(2)$ group with Pauli
 matrices $T^a=\sigma^a/2$, we have
\begin{equation}
Q^1=\frac{J(\xi+{\bar \xi})}{1+\mid\xi\mid^2},\ \ \ \ Q^2=i\frac{J
({\bar \xi}-\xi)}{1+\mid\xi\mid^2},\ \ \ \
Q^3=\frac{J(1-\mid\xi\mid^2)}{1+\mid\xi\mid^2}
\end{equation}
In terms of stereographical projection $\xi=\tan (\theta/2)e^{i\phi}$,
 they are just
the ordinary angular momentum :
\begin{equation}
Q^1= J \sin \theta \cos\phi,\quad Q^2=
J\sin \theta \sin \phi,\quad Q^3 =
J\cos\theta
\end{equation}

To calculate the Poisson brackets for the momentum map function $Q^a$
defined by Eq.(\ref{poi}),
we must use the symplectic structure on ${\bf CP}^{n}$. To do that, we
introduce the notation $\xi^A=(\bar\xi_p, \xi_q)$ and write the symplectic
two form as $\Omega=\frac{1}{2}\Omega_{AB}d\xi^A\wedge d\xi^B$.
The Poisson bracket defined by Eq.(\ref{poi})
 with the use of Eq.(\ref{omeg})
have the following expression:
\begin{equation}
\{F,H\}=-i\sum_{p,q}g^{pq}\left(\frac{\partial
F}{\partial {\bar \xi}_p}\frac{\partial H}{\partial \xi_q}-
\frac{\partial F}{\partial\xi_q}\frac{\partial H}{\partial {\bar
\xi}_p}\right),
\end{equation}
where $g^{pq}$ is the inverse of the Fubini-Study metric given by
\begin{equation}
g^{pq}=\frac{1}{2J}(1+\mid\xi\mid^2)
(\delta_{pq}+{\bar \xi}_p\xi_q).\label{metric}
\end{equation}
A simple calculation yields the fundamental commutators as follows
\begin{equation}
\{{\bar \xi}_p,\xi_q\}=-\frac{i}{2J}(1+\mid\xi\mid^2)
(\delta_{pq}+{\bar \xi}_p\xi_q),
\label{poi3}
\end{equation}
\[\{\xi_p,\xi_q\}=\{{\bar \xi}_p,{\bar \xi}_q\}=0,\]
and shows that the momentum map functions $Q^a$
satisfy $su(n+1)$ algebra
\begin{equation}
\{Q^a,Q^b\}=-f^{abc}Q^c.\label{op}
\end{equation}

\section{Associated Bundle and Symplectic Structure}
\setcounter{equation}{0}
Consider two dimensional configuration space $M$ which
could be in general arbitrary Riemann surfaces. For simplicity, we assume
that $M$ is a plane.
(We present one particle case and later extend to $N$
particles in a straightforward manner.)
 Let ${\bar P}\rightarrow M$ be a principal $G$ bundle over $M$.
Let $X=T^*M$ and $P$ be the pull-back of the bundle  ${\bar P}$
by the projection $\pi^\prime:T^*M\rightarrow M$.
Then $P$ is a principal $G$ bundle over $X$.
Because $P$ is a principal $G$ bundle and $G$ acts on ${\cal O}$, we
can form the associated bundle ${\cal P}\equiv P\times_G{\cal O}$.
${\cal P}$ is the phase space of an
isospin particle under the influence of external gauge field.
 Sternberg showed\cite{ster} that given a connection $\Theta$ on $P$,
there exists a unique symplectic structure defined on ${\cal P}$. We
will explicitly calculate this symplectic structure for ${\cal O}=
{\bf CP}^{n}$.

We present the essence of Sternberg's results. Let $\zeta_P$ denote
vector field on $P$ generated by the one parameter group consisting
of right multiplication by $\exp(-t\zeta):\zeta_P(p)=
d/dt{\big\vert}_{t=0}p\circ \exp(-t\zeta), p\in P$. Let $R_g$ denote the
right multiplication $R_g(p)=pg^{-1}$. The connection $\Theta$
on $P$ is defined as a ${\cal G}$-valued differential one form
which satisfies the following two conditions:
\begin{equation}
R_g^*\Theta=Ad_g\Theta, \quad \Theta(\zeta_P)=\zeta
\end{equation}
for all $\zeta\in {\cal G}$.
The infinitesimal version of first condition is
${\cal L}_{\zeta_P}\Theta=ad_\zeta\Theta$.
 Let $U_g$ denote the action of $g\in G$ on
$ P\times{\cal O}$ by $U_g(p,f)=(pg^{-1},gf)$.

By definition, the associated
bundle  ${\cal P}$ is the quotient space of  $ P\times{\cal O}$
 under the action
of $U_g$.
Now define the real valued differential one form
$\langle \Theta, Q \rangle$ on
$ P\times{\cal O}$. Then one can easily show
 that $\langle \Theta, Q \rangle$
is invariant under the action of $U_g:U_g^*\langle \Theta, Q \rangle=
\langle \Theta, Q \rangle$. However, this does not imply that
$\langle \Theta, Q \rangle$ is well defined on the quotient space ${\cal P}$.
One can show this by proving that $d\langle \Theta, Q \rangle$ is not
defined on ${\cal P}$.  A simple calculation
using ${\cal L}_{\zeta_P}\Theta=ad_\zeta\Theta$
indeed shows that
\begin{equation}
\zeta_U\rfloor d\langle \Theta, Q \rangle
=-\pi^*(\zeta_{{\cal O}}\rfloor \Omega).
\label{stern}
\end{equation}
Here $\zeta_U$ denotes the vector field on $P\times {\cal O}$
 corresponding
to the action $U$ on $P\times {\cal O}$
so that $\zeta_U=\zeta_P+\zeta_{{\cal O}}$ and $\pi$ is the
projection map $\pi: P\times{\cal O}\rightarrow {\cal O}$.
 The above equation implies
that $d\langle \Theta, Q \rangle$ is not naturally defined
on the quotient space
${\cal P}$ because it does not vanish when evaluated on vectors, one of
which is along the direction of projection
 $P\times{\cal O}\rightarrow {\cal P}$.

To have expression for two form which descends on ${\cal P}$, consider
$d\langle \Theta, Q \rangle+\pi^* \Omega$.
Then, we have
\begin{equation}
\zeta_U\rfloor (d\langle \Theta, Q \rangle+\pi^* \Omega)=0.
\end{equation}
by using  Eq.(\ref{stern}). One can also easily check that
$d\langle \Theta, Q \rangle+\pi^* \Omega$
 is invariat under the $U_g$ action.
These suggest that there exist a unique form $\Omega_{\Theta}$
on ${\cal P}$ such that
\begin{equation}
d\langle \Theta, Q \rangle+\pi^* \Omega=\bar\pi^*\Omega_{\Theta}
\end{equation}
where $\bar\pi$ is the projection map
$\bar\pi:P\times{\cal O}\rightarrow {\cal P}$.
Since $\bar\pi^*$ is one-to-one and
 $d\langle \Theta, Q \rangle+\pi^* \Omega$
is closed, the two form
 $\Omega_{\Theta}$ is closed:$d\Omega_{\Theta}=0.
$ Also, $\Omega_{\Theta}$is nondegenerate in the case
when $X=T^*M$ and $P$ is the pull-back of the bundle  ${\bar P}$
and the connection is the pull-back of a connection defined on ${\bar P}$
\cite{ster}.
Denoting  $\tilde\omega$ for the pull-back of $\omega$ which is the
canonical symplectic structure defined on $X$  to ${\cal P}$
via projection onto $X$, we have symplectic structure on ${\cal P}$ as
\begin{equation}
\Omega_T=\tilde\omega+\Omega_{\Theta}
\end{equation}

When $M$ is a plane, $T^*M$ is contractible and every associated bundle
is trivial. So we have
\begin{equation}
{\cal P}= P\times_G{\cal O}=T^*M\times {\cal O}
\end{equation}
In fact, the above holds for arbitrary Riemann surfaces $M$. This  can
be seen from the following simple homotopy arguement.
It is well known that the set of equivalence classes
of principal G-bundles $K_G(T^*M)$ is isomorphic to
$[T^*M,BG]$  where $[T^*M,BG]$ is the set of homotopy
classes of maps from
$T^*M$ to the classifying space $BG$ of $G$.
Consider the homotopy exact sequence of universal principal $G$-bundle
\begin{equation}
\pi_i(G)\rightarrow \pi_i(EG)\rightarrow \pi_i(BG)\rightarrow
\pi_{i-1}(G)\rightarrow \pi_{i-1}(EG).
\end{equation}
Since $EG$ is contractible, $\pi_i(EG)=\pi_{i-1}(EG)=0$ and  we have
$\pi_i(BG)= \pi_{i-1}(G)$. By the fact\cite{huse} that
\begin{equation}
\pi_k(SU(n+1))=\left\{ \begin{array}{ll}
			   0    &    \{k\leq 2\}\\
		     {\bf Z}    &    \{k=3 \},\end{array}
			   \right.
\end{equation}
we have
\begin{equation}
 \pi_k(BG)=\left\{ \begin{array}{ll}
			   0    &    \{k\leq 3\}\\
		     {\bf Z}    &    \{k=4 \}\end{array}
			   \right. .
\end{equation}
Consider the reduced singular cohomology
group $\tilde H^k(T^*M,\pi_k(BG))$. Since $T^*M$ is homotopy
equivalent to $M$, $\tilde H^k(T^*M,\pi_k(BG))=\tilde H^k(M,\pi_k(BG))$.
Since $M$ is a real two dimensional manifold, $\tilde H^k(M,\pi_k(BG))=0$
for $k\geq 3$. If $k=1$, $\pi_1(BG)=0$ and $\tilde H^1(M,\pi_1(BG))=0.$
Thus there exists
a surjective map\cite{gray}  from $\tilde H^2(T^*M,\pi_2(BG))=
 \tilde H^2(M,\pi_2(BG))$ to $[T^*M,BG]$.
Since for $k=2$, $\pi_2(BG)=0$ and $\tilde H^2(M,\pi_2(BG))=0$,
 this implies that every principal fiber bundle over $M$ ia trivial
and we have ${\cal P}= P\times_G{\cal O}=T^*M\times {\cal O}$.
Hence we have $\tilde\omega=\omega$ and
\begin{eqnarray}
\Omega_T & = & \omega+
\sigma^*(d\langle \Theta, Q \rangle+\pi^* \Omega) \nonumber  \\
      & = & \omega +d(A^aQ^a)+\Omega
\end{eqnarray}
where $\sigma$ is the cross section :$ P\times_G{\cal O}\rightarrow
 P\times{\cal O}$ and
we used $\sigma^*\Theta=A$, the gauge field one form on $M$.
 Notice that $\omega+\Omega$ is not
gauge invariant. We must have Sternberg's two form
 $d\langle \Theta, Q \rangle$
to achieve the gauge invariance.
 Physically, this term describes the interaction
between isospin cahrge and the external gauge fields.
 Now, we calculate the
symplectic structure on $P\times_G{\cal O}$.
We start from the two form on $P\times_G{\cal O}=
T^*M\times {\cal O}$ given by
\begin{equation}
\Omega_T=dp_i\wedge dq^i+ d(A^a_i Q^a dq^i) +\Omega
\end{equation}
where the $\Omega$ is given by the expression Eq.(\ref{omeg}).
 To achieve the
notational simplifications, we introduce $\eta^I=(p_i, q^j)$ and
$x^M=(\xi^A,\eta^I)$. Then we can write $\Omega_T=
\frac{1}{2}\Omega_{MN}dx^M\wedge dx^N$ where the matrix
 $\Omega_{MN}$ can be expressed as
\begin{equation}
\Omega_{MN}=\left( \begin{array}{cc}
   \Omega_{AB} & A^a_J(\partial Q^a/\partial \xi^A) \\
  -A^a_I(\partial Q^a/\partial \xi^B) & \omega_{IJ}
	\end{array}  \right).\label{def1}
\end{equation}
Here, $A^a_I=(0,A^a_i)$ and $\omega_{IJ}$ is given by
\begin{equation}
\omega_{IJ}=\left(\begin{array}{cc}
	0 & I \\
       -I & A^a_{[j,i]}Q^a \end{array} \right).\label{def3}
\end{equation}
A short calculation gives the following inverse matrix $\Omega^{MN}$:
\begin{equation}
\Omega^{MN}=\left( \begin{array}{cc}
    \Omega^{AB}  & -F^{KJ}\Omega^{AC}A^a_K(\partial Q^a/\partial \xi^C) \\
F^{KI}\Omega^{BD}A^a_K(\partial Q^a/\partial\xi^D) & F^{IJ}
\end{array} \right)
\label{mini1}
\end{equation}
where $F^{IJ}$ is the inverse matrix of
 $F_{IJ}\equiv \omega_{IJ}-f^{abc}A^a_IA^b_JQ^c$.
Using Eq.(\ref{def3}), we find the matrix  $F^{IJ}$:
\begin{equation}
 F^{IJ}=\left(\begin{array}{cc}
   F^a_{ij}Q^a & -I  \\
	     I & 0 \end{array} \right).
\label{def4}
\end{equation}
where $F^a_{ij}\equiv \partial_jA^a_i-\partial_iA^a_j-f^{abc}A^b_iA^c_j$
is the Yang-Mills field strength.

The Poisson bracket on $P\times_G{\cal O}$ is defined by the use of
inverse matrix $\Omega^{MN}$ as before
\begin{equation}
\{F,H\}=\Omega^{MN}\frac{\partial F}{\partial x^M}
\frac{\partial H}{\partial x^N}.
\end{equation}
The calculation  is greatly simplified by the use
of Eq.(\ref{op})
and we find the following Poisson bracket among the dynamical
 variables
\begin{equation}
\{Q^a,p_i\}=-f^{abc}A^b_iQ^c, \ \ \ \ \ \ \ \ \{Q^a,q^i\}=0
\end{equation}
\[ \{p_i, p_j\}=F^a_{ij}Q^a, \ \ \ \{p_i,q^j\}=-\delta_i^j,
 \ \ \ \{q^i,q^j\}=0.\]
The above relations are in accordance with the minimal substituion
\begin{equation}
p_i\rightarrow P_i=p_i - A_i^a Q^a.
\end{equation}
In terms of canonical momentum $P_i$, we have, among others,
\begin{equation}
\{Q^a,P_i\}=0 \ \ \ \ \{P_i, P_j\}=0. \ \ \ \ \{P_i,q^j\}=-\delta_i^j.
\label{cano}
\end{equation}
Thus, one can work in $(p_i,q^i,Q^a)$
coordinates using the symplectic
structure given by Eqs.(\ref{mini1}) and (\ref{def4}) or with
 $(P_i,q^i,Q^a)$ using the
canonical symplectic structure without mixing between
 $ P_i$ and $Q^a$.
The two procedures are equivalent\cite{ster}.
Consider, for example, the free Hamiltonian $H=(1/2m) p^2$
 with symplectic
structure given by Eqs.(\ref{mini1}) and (\ref{def4}).
 The Hamiltonian equations of motion
\begin{equation}
\dot x^M=\Omega^{MN}\frac{\partial H}{\partial x^N}
\end{equation}
reproduces the well known Wong's equations
\begin{equation}
m\ddot{q}_i =  F^a_{ij}Q^a\dot{q}^j \quad
\dot{Q}^a = -f^{abc} A^b_i \dot{q}^i Q^c ,\label{eul2}
\end{equation}
which describes the dynamics of a isospin particle in external
gauge fields $A_i^a$.
Minimal substitution implies that alternatively, we can work with
\begin{equation}
 H =  {1\over 2 m}\left(P_i-A^a_i Q^a\right)^2,\label{hamil}
\end{equation}
with canonical symplectic structure Eq.(\ref{cano}).
Obviously, we get the same equations of motions.
The above procedures can be generalized to a system of $N$ particles
in a obvious manner. We will
consider from now on a system of $N$ particles and
denote the particle index
by $\alpha,\beta,\cdots=1,\cdots,N$.

\section{Reduced Phase Space and Connection}
\setcounter{equation}{0}
The analysis we have done so far holds for most part in describing isospin
particles under the influence of arbitrary external gauge fields.
In this section, we attempt to determine a specific connection which is
essential in describing the quantum mechanics of fractional spin and
braid statistics.
To do that,  we first consider phase space
${\cal P}_T\equiv \prod_\alpha
 {\cal P}^\alpha\times {\cal A}=\prod_\alpha
T^*M^\alpha\times {\cal O}^\alpha
\times {\cal A}$ where ${\cal A}$ is the space of
all connections.
We define the  gauge transformation of
 $f\in {\cal O}$ and $A\equiv A^a_iT^a dq^i \in {\cal A}$ by
\begin{equation}
f^g_\alpha=gf_\alpha, \quad A^g= g^{-1}Ag+ig^{-1}dg \label{gt}
\end{equation}
for $g\in G=SU(n+1)$. The momentum map
 for $G$ action on $\prod_\alpha
{\cal O}^\alpha=\prod_\alpha C{\bf P}^{\alpha n}$ is
$Q= \sum_{\alpha}Q^a_\alpha
 T^a\delta({\bf x}-{\bf q_\alpha})dx^1\wedge dx^2$
with $Q^a$ given by Eq.(\ref{def2}). To describe the momentum map for
$G$ action on  ${\cal A}$, consider the symplectic two form on  ${\cal A}$:
\begin{equation}
\Omega_{ {\cal A}}(a,b)=\kappa\int_{M} Tr(a\wedge b)
\end{equation}
where $a,b\in {\cal G}$ are tangent vectors at $A\in {\cal A}$ and
one form on $M$. $\kappa$ is the coefficient of the Chern-Simons gauge
theory. The above
form is invariant under the $G$ action (\ref{gt}) and the corresponding
momentum map is given by the curvature two form $\kappa F$,
$F=dA+iA\wedge A\equiv (1/2)F_{ij}^aT^a dx^i\wedge dx^j$\cite{atiy}.
To see this first define the
momentum map function $\mu_\zeta $ on $ {\cal A}$ by
\begin{equation}
\mu_\zeta (A)\equiv \langle \zeta,\mu_A\rangle= \int_{M}
 Tr(\mu_A\wedge \zeta)
\end{equation}
where $\zeta\in {\cal G}$ are zero forms on $M$.
 The above definition suggests
that the momentum map is two form on $M$.  The vector field generated by
$\zeta$ is given by the gauge
 transformations $d_A\zeta$ and from the definition
of momentum map Eq.(\ref{moment}), we have for arbitrary $a$
\begin{equation}
\langle a, d\mu_\zeta(A)\rangle=
d_A\zeta\rfloor \Omega_{{\cal A}}(a)=\kappa\int_{M}
Tr(d_A\zeta\wedge a)
=-\kappa\int_{M} Tr(\zeta\wedge d_Aa)
\end{equation}
where we  integrated by parts in the last step.
We see that the last term is the gauge
 transformation of $F_\zeta$ along the
 direction of $a$ and can be written as $\kappa\langle a,
dF_\zeta(A) \rangle$\cite{atiy}.
So we have $\mu_A=\kappa F$.
Hence the total momentum map is given by
 $\Phi=Q+\mu_A=Q+\kappa F$ .
We can form the quotient space $\Phi^{-1}(0)/G$ which is the
Marsden-Weinstein reduction\cite{guil}
and this moduli space is the reduced phase space
of the NACS particles. Note that the  $\Phi=0$ can be written as
\begin{equation}
\Phi^a = {\kappa \over 2}\epsilon^{ij} F^a_{ij} ({\bf x}) +
\sum_\alpha Q^a_\alpha \delta({\bf x}- {\bf q}_\alpha) =
0,\label{gaus}
\end{equation}
which is the Gauss constraints in Chern-Simon theory coupled with
point sources.

Rather than probing the geometry of the reduced phase space,
 we attempt to
determine a specific connection as the solution of
the above Gauss constraints which can be solved  explicitly
in two gauge conditions.
The first one is the axial gauge\cite{kapu}
in which, for example, we set $A_1^a=0.$  The remaining $A_2^a$ field
becomes highly singular with strings attached to each source
and we do not adopt this solution.
The less singular solutions can be obtained by performing
the analytic continuation of the gauge fields.
Introducing complex coordinates, $z = x+ iy$, $\bar z
= x- iy$, $z_\alpha = q^1_\alpha + i q^2_\alpha$, $\bar z_\alpha =
q^1_\alpha
- i q^2_\alpha$, $A^a_z = {1\over 2} (A^a_1 - iA^a_2)$, $A^a_{\bar z} =
{1\over 2} (A^a_1 + iA^a_2)$, analytic continuation means that
$A^a_z$ and $A^a_{\bar z}$ are treated as independent variables.
We recall\cite{dani} that choosing a gauge condition
corresponds to reduction  to the space $\Phi^{-1}(0)/G$ by the choice of
representatives in $\Phi^{-1}(0)$ of all orbits. In the analytic continuation,
the representatives are fixed  in the complexified guage orbit space and
this is consistent with the coherent state quantization method\cite{fadd2}.
We  choose $A^a_{\bar z}=0$ as a gauge fixing condition
in this space which was called holomorphic gauge in ref\cite{lo1}.
The solution of the Gauss constraints
\begin{equation}
\Phi^a(z)=-\kappa \partial_{\bar z} A^a_z +\sum_\alpha
Q^a_\alpha\delta(z-z_\alpha)=0, \label{gauss}
\end{equation}
in holomorphic gauge turns out to be\cite{lo1}
\begin{equation}
A^a_z (z, \bar z) = {i\over 2\pi \kappa}\sum_\alpha  Q^a_\alpha
{1\over z -z_\alpha}+P(z),\label{sol}
\end{equation}
where $P(z)$ is an arbitrary holomorphic polynomial in $z$.
The further choice of $P(z)=0$ resulted in the quantum mechanical model
which  provides a unified framework for fractional spin, braid statistics and
Knizhnik-Zamolodchikov equation\cite{kniz}.
In other words, essential ingredients in the quantum mechanical model of
Ref.\cite{lo1}
consist of classical phase which is  $\prod_\alpha
T^*M^\alpha\times {\cal O}^\alpha$ and a specific connectionin of
Eq.(\ref{sol})
with $P(z)=0$. This connection known as Knizhnik-Zamolodchikov (KZ)
connection in the literature
plays an  important role in establishing the relation of Chern-Simons theory
with conformal field theory and quantum group \cite{khon,guad}.
Substituting the above connection into the $N$ particle Hamiltonian where
Hamiltonian of each particle is given by Eq.(\ref{hamil}),
 we obtain(replacing $P_i\rightarrow p_i$)
\begin{equation}
H=\sum_\alpha \frac{2}{m_\alpha}p_\alpha^z\left(p^{\bar z}_\alpha-
{i\over 2\pi \kappa}\sum_\beta \frac{ Q^a_\alpha  Q^a_\beta}
{ z_\alpha -z_\beta}\right)\label{prop}
\end{equation}
Quantum mechanically, the dynamics of the NACS particles are
 governed by the
operator version  ${\hat H}$ of the Hamiltonian Eq.(\ref{prop})
\[ \hat {H} = -\sum_\alpha {1\over m_\alpha}\left(\nabla_{\bar
z_\alpha}\nabla_{z_\alpha}  +\nabla_{z_\alpha}\nabla_{\bar
z_\alpha}\right) \]
\begin{equation}
\nabla_{z_\alpha} ={\partial\over \partial z_\alpha}  +{1\over 2\pi
\kappa}\left( \sum_{\beta\not=\alpha}
\hat Q^a_\alpha \hat Q^a_\beta {1\over
z_\alpha -z_\beta}+\hat Q^2_\alpha a_z (z_\alpha)\right)\label{ham}
\end{equation}
\[\nabla_{\bar z_\alpha} ={\partial\over \partial \bar z_\alpha}\]
where $a_z (z_\alpha)=\lim_{z\rightarrow z_\alpha} 1/(z-z_\alpha)$
and the
isospin operators $\hat Q^a$'s satisfy the $SU(N+1)$
algebra, $[\hat Q^a_\alpha,\hat Q^b_\beta] =if^{abc} \hat
Q^c_\alpha \delta_{\alpha\beta}$ upon quantizing the classical algebra
Eq.(\ref{op}). The second term and the third term
in $\nabla_{z_\alpha}$ are
reponsible for the non-Abelian statistics and the the
anomalous spins of the NACS particles respectively.
The detailed analysis of  the above quantum mechanical model
has been performed in  Ref.\cite{lo1,lee}.

\section{Conclusion}
\setcounter{equation}{0}
In this paper, we have investigated the phase space structure of $N$ NACS
particles from a geometrical point of view. We first considered the product
space of $N$ associated $SU(n+1)$ bundle with ${\bf CP}^n$ as the fiber.
 The  momentum map which is the essential ingredient necessary to probe
the phase space structure and to construct the gauge invariant symplectic
two form was obtained for ${\bf CP}^n$. An interesting consequence of
a simple
homotopy arguement was that the associated bundle is equivalent to the
direct product of canonical cotangent bundle by the ${\bf CP}^n$ manifold
for arbitrary two-dimensional Riemann surfaces.
The explicit evaluation of the symplectic structure showed that
the minimal substitution of canonical momentum in the external non-Abelian
gauge fields is equivalent to the modification of canonical
 symplectic structure
with the addition of field strength two form as in the Abelian case.
Then, we introduced the space of all Chern-Simons connections
 and adopted the
procedure of symplectic reduction in order to obtain
 the reduced phase space of
NACS particles.  A specific connection was determined by
the condition of vanishing momentum map
with the holomorphic gauge choice
and this produced the well-known KZ connection.
 This connection endows NACS particles with the
 non-Abelian magnetic flux
and introduces topological interaction between them.
The canonical quantization for two particle system
using this connection was performed\cite{lee} in detail.
The geometric quantization of such a model is
 another interesting possibility
and will be reported elsewhere.

\acknowledgments

One of us(PO) would like to thank Prof. T. Lee for useful discussions.
This work was supported in part by  the KOSEF through C.T.P. at S.N.U.
and by the  Ministry of Education through the Research Institute of
Basic Science.


\begin{references}
\bibitem[a)]{mhkim} E-mail address:mhkim@yurim.skku.ac.kr
\bibitem[b)]{poh} E-mail address:ploh@yurim.skku.ac.kr
\bibitem{lo1} T. Lee and P. Oh, Phys. Rev. Lett. {\bf 72}, 1141 (1994).
\bibitem{wilz3} F. Wilczek, ed., {\it Fractional Statistics and
 Anyon Superconductivity} (World Scientific, Singapore, 1990).
\bibitem{fort} S. Forte, Rev. Mod. Phys. {\bf 64}, 193 (1992).
\bibitem{ieng} R. Iengo and K. Lechner, Phys. Rep. {\bf 213}, 179 (1992).
\bibitem{pran} R. E. Prange and S. M. Girvin, ed.,
 {\it The Quantum Hall Effect} (Springer, Berlin, 1990);
M. Stone, ed., {\it Quantum Hall Effect}  (World Scientific, Singapore, 1992).
\bibitem{dese} S. Deser, R. Jackiw and S. Templeton, Phys. Rev. Lett.
{\bf 48}, 975 (1982); Ann. Phys. (N.Y.) {\bf 140}, 372 (1982).
\bibitem{witt} E. Witten, Commun. Math. Phys. {\bf 121}, 351 (1989).
\bibitem{bala} A. P. Balachandran, M. Bourdeau  and S. Jo,  Mod.
 Phys. Lett. A {\bf 4},1923 (1989) ; Int. J. Mod. Phys. A {\bf 5}, 2423 (1990).
\bibitem{verl} E. Verlinde, in {\it Modern Quantum Field Theory} (World
		     Scientific, Singapore, 1991).
\bibitem{froh} J. Fr\"ohlich, in {\it Non-Perturbative Quantum Field
Theory} edited by G. 't Hooft {\it et al.} (Plenum, New York, 1988).
\bibitem{oh} T. Lee and P. Oh, Phys. Lett. B {\bf 319}, 497 (1993).
\bibitem{bak} D. Bak, R. Jackiw and S.-Y. Pi, Preprint  MIT
CTP \#  2276, hep-th/9402057.
\bibitem{kiml} W. T. Kim and C. Lee, SNUTP 94-14.
\bibitem{wein} A. Weinstein, Lett. Math. Phys. {\bf 2}, 417 (1978).
\bibitem{ster} S. Sternberg, Proc. Nat. Acad. Sci. U.S.A. {\bf 74}, 5253
 (1977); V. Guillemin and S. Sternberg, Hadronic J. {\bf 1}, 1 (1978).
\bibitem{kril} A. A. Kirillov, {\it Elements of the Theory of
 Representations} (Springer-Verlag, 1976).
\bibitem{wong} S. K. Wong, Nuovo Cimento {\bf 65A}, 689 (1970).
\bibitem{guil} V. Guillemin and S. Sternberg, {\it Symplectic Techniques
in Physics} (Cambridge Univ., Cambridge, 1984).
\bibitem{bere} F. A. Berezin, Comm. Math. Phys. {\bf 40}, 153 (1975);
J. R. Klauder, Phys. Rev. D {\bf 19}, 2349 (1979);
E. Witten, Comm. Math. Phys. {\bf 92}, 455 (1984).
\bibitem{fa} L. Faddeev and R. Jackiw, Phys. Rev. Lett. {\bf 60}, 1692 (1988).
\bibitem{grif} P. Griffiths and J. Harris, {\it Principles of Algebraic
Geometry} (Wiley, New York, 1978).
\bibitem{huse} D. Husemoller, {\it Fiber Bundles}, 2nd Edition
(Springer-Verlag, 1966).
\bibitem{gray} B. Gray, {\it Homotopy Theory: An Introduction to
Algebraic Topology} (Academic Press, 1975).
\bibitem{atiy} M. F. Atiyah and R. Bott, Phil. Trans. R. Soc. London A
{\bf 308}, 523 (1982).
\bibitem{kapu} A. N. Kapustin and P. I. Pronin, Phys. Lett.
 B {\bf 303}, 45 (1993).
\bibitem{dani} M. Daniel and C. M. Viallet, Rev. Mod. Phys. {\bf 52},
175 (1980).
\bibitem{fadd2} L. D. Faddeev and A. A. Slavnov, {\it Gauge Fields:
Introduction to Quantum Theory} (Benjamin/Cummings Pub., MA, 1980);
 L. S. Brown, {\it Quantum Field Theory} (Cambridge
Univ. Press, 1992).
\bibitem{kniz} V. G. Knizhnik and A. B. Zamolodchikov, Nucl. Phys. B
{\bf 247}, 83 (1984).
\bibitem{khon} T. Khono, Ann. Inst. Fourier {\bf 37},
		139 (1987).
\bibitem{guad} E. Guadagnini, M. Martellini and M. Mintchev, Nucl. Phys.
  B {\bf 336}, 581 (1990).
\bibitem{lee} T. Lee and P. Oh, SNUTP-93-82 ,
 Ann. Phys. (N.Y.) in press.

\end{references}
\end{document}